\documentclass[prb,floatfix,twocolumn,showpacs,aps]{revtex4}%
\usepackage{amsfonts}
\usepackage{amsmath}
\usepackage{amssymb}
\usepackage{graphicx}%
\begin{document}
\title{A quest for frustration driven distortion in Y$_{2}$Mo$_{2}$O$_{7}$}
\author{Eva Sagi, Oren Ofer, and Amit Keren}
\affiliation{Physics Department, Technion, Israel Institute of Technology, Haifa 32000, Israel}
\author{Jason S. Gardner}
\affiliation{Physics Department, Brookhaven National Laboratory, Upton, New York 11973-5000
\& NIST Center for Neutron Research, National Institute of Standards and
Technology, Gaithersburg, Maryland 20899-8562}

\begin{abstract}
We investigated the nature of the freezing in the geometrically frustrated
Heisenberg spin-glass Y$_{2}$Mo$_{2}$O$_{7}$ by measuring the temperature
dependence of the static internal magnetic field distribution above the
spin-glass temperature, $T_{g}$, using the $\mu$SR technique. The evolution of
the field distribution cannot be explained by changes in the spin
susceptibility alone and suggests a lattice deformation. This possibility is
addressed by numerical simulations of the Heisenberg Hamiltonian with
magneto-elastic coupling at $T>0$.

\end{abstract}
\date[Date text]{date}
\maketitle

A geometrically frustrated magnet is one whose symmetry precludes every
pairwise interaction in the system being satisfied simultaneously. Such
systems can have \textquotedblleft macroscopic\textquotedblright\ degenerate
ground states, which prevents the system from settling into a unique state as
the temperature is lowered. In this situation, small perturbations to the
Hamiltonian can have a dramatic effect, selecting one ground state over
another. It has been suggested that the ground state degeneracy of the
Heisenberg antiferromagnet on the pyrochlore lattice, a lattice of
corner-sharing tetrahedra, can be lifted by a magneto-elastic coupling
\cite{KTerao,YamashitaUeda,OlegPRB,RichterPRL04,KerenPRL,Evidence}. Such a
coupling allows the lattice to distort in order to relieve the magnetic
frustration, concomitantly lowering the total system energy. Since the 3D
corner-sharing tetrahedra lattice is susceptible to distortion under
arbitrarily small magneto-elastic coupling, it is at the centre of an ongoing
quest to observe this phenomena. Several materials show a lot of promise in
this respect, including several Cr$^{3+}$ spinels and several A$_{2}$B$_{2}%
$O$_{7}$ pyrochlores \cite{KerenPRL,Evidence,Sushkov}.

One promising, but puzzling, candidate is Y$_{2}$Mo$_{2}$O$_{7}$ [YMO]. YMO is
a narrow-band semiconductor with a Curie-Weiss temperature of $\sim$ 200~K and
a spin-glass transition at $22.5$~K \cite{Raju}. The glassiness of the system
implies some kind of disorder in the exchange integral, whose origin might be
a deformed lattice. Indeed, Booth \textit{et al.} \cite{Booth} showed, using
the X-ray-absorption fine-structure technique, that the Mo tetrahedra are in
fact distorted at the local level by about 5\%. This relatively large amount
of bond disorder is not seen by the usual diffraction techniques (X-rays or
neutrons), indicating that the average bulk structure is almost the perfect
oxide pyrochlore lattice \cite{Raju}. Further evidence of disorder was
revealed by $^{89}$Y NMR \cite{KerenPRL}, where regularly spaced peaks implied
the existence of many non-equivalent $^{89}$Y sites. These features were
attributed to distortions in the Mo-sublattice. However, the NMR data were
limited to nitrogen temperatures due to limitations on the maximum RF power
available in Helium atmosphere, and the decrease of signal intensity with
decreasing temperature due to the abnormal increase of the line width. To gain
a better understanding of the freezing process in YMO, and to determine
whether a lattice deformation is active close to $T_{g}$, we performed a study
similar to the NMR one and with the same sample, using muon spin rotation and
relaxation techniques. In this letter we present these $\mu$SR results and
attempt to reconcile them (and the NMR) with the existing diffraction data
through numerical simulations.

\begin{figure}
[ptb]
\begin{center}
\includegraphics[
natheight=10.849900in,
natwidth=8.551300in,
height=3.0502in,
width=2.4094in
]%
{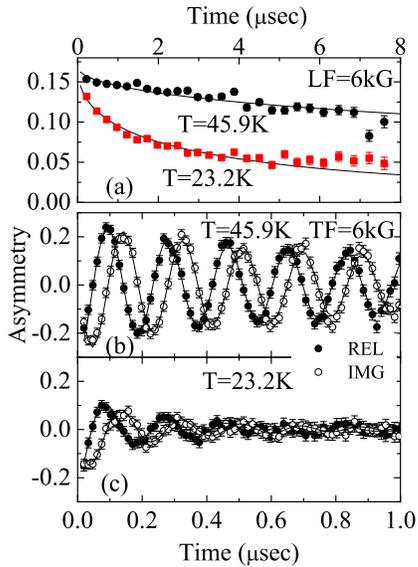}%
\caption{(a) Longitudinal field asymmetries for $T=23.2$ K and $T=45.2$~K; (b)
and (c) real and imaginary transverse field asymmetries for $T=45.2$~K and
$T=23.2$ K respectivly. The solid lines are fits to Eqs~\ref{LFFit} and
\ref{TFFit}.}%
\label{RawData}%
\end{center}
\end{figure}

Transverse [TF] and longitudinal field [LF] $\mu$SR measurements were
performed on the GPS spectrometer at the Paul Scherrer Institute, Switzerland.
The measurements were carried out with the muon spin tilted by $50^{0}$
relative to the direction of the applied magnetic field of 6~kG, and positron
data were accumulated in both the forward-backward (longitudinal) and the
up-down (transverse) directions simultaneously. This allows us to determine
both transverse and longitudinal muon spin relaxations rates. In Fig.
\ref{RawData} we show the LF [panel (a)] and the TF data [panels (b) and (c)]
at two temperatures. Both TF and LF relaxation rates increase with decreasing
temperature; however, at 23.2~K the transverse asymmetry is zero after 0.6
$\mu$sec while the longitudinal asymmetry is still finite at much longer
times. This means that close to $T_{g}$ the TF relaxation is larger than the
LF one and most of the in-plane depolarization is from static field inhomogeneities.%

We found that the $\mu$SR LF asymmetry is best described by the root
exponential
\begin{equation}
A_{LF}(t)=A_{0}\exp(-(R_{LF}t)^{\frac{1}{2}})+B_{g} \label{LFFit}%
\end{equation}
where the parameters $A_{0}$ are set by taking into account the tilt of the
muon spin relative to the longitudinal magnetic field, $R_{LF}$ is the
longitudinal relaxation rate, and $t$ is time. Similarly, the TF asymmetry is
best fitted by a root exponential superimposed on a cosine oscillation with a
transverse relaxation rate $R_{TF}$:
\begin{equation}
A_{TF}(t)=A_{0}\exp(-(R_{TF}t)^{\frac{1}{2}})cos(\omega t+\phi)+B_{g}.
\label{TFFit}%
\end{equation}
$A_{0}$, the muon rotational frequency ($\omega=\gamma_{\mu}H_{TF}$), and the
background ($B_{g}$) are common for all the TF data sets. Since the TF
relaxation is a result of both static field inhomogeneity and dynamically
fluctuating fields, and the LF relaxation stems from dynamic fluctuations
only, we can extract the rates of the two processes. The static field
distribution alone will lead to a polarization function of the form
\begin{equation}
P_{static}(t)=P_{0}\exp(-(\Delta t)^{1/2})cos(\omega t)
\label{polarization_root}%
\end{equation}
where $\Delta=[R_{TF}^{1/2}-R_{LF}^{1/2}]^{2}$ assuming isotropic fluctuations.

\begin{figure}
[ptb]
\begin{center}
\includegraphics[
natheight=8.754500in,
natwidth=11.251200in,
height=2.4881in,
width=3.1912in
]%
{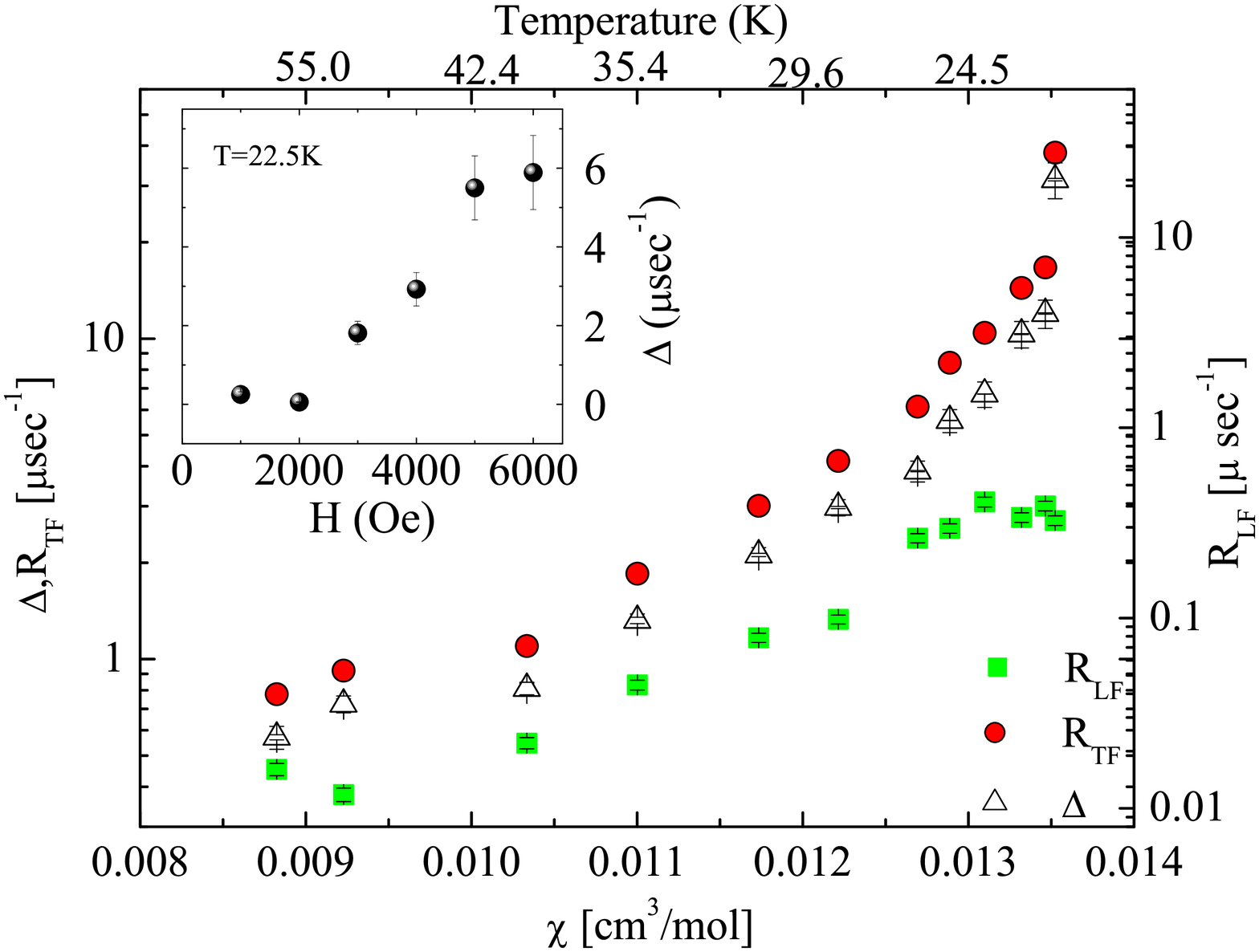}%
\caption{A plot of the static ($\Delta$)$,$ longitudinal (R$_{LF}$) and
transverse (R$_{TF}$) relaxations rates, on a log scale, versus susceptibility
and temperature. In the inset we present $\Delta$ versus magnetic field at a
constant temperature of $22.5$~K.}%
\label{T22p5}%
\end{center}
\end{figure}

In Fig. \ref{T22p5} we plot $R_{TF}$, $R_{LF}$, and $\Delta$ on a log scale
vs. $T$ on the upper part of the abscissa. As the temperature is lowered the
difference between $R_{TF}$ and $R_{LF}$ increases. In addition $\Delta$ grows
exponentially on this semi-log plot upon cooling. In the inset of Fig.
\ref{T22p5} we present $\Delta$ for various fields at $T=22.5$~K. One can
clearly see that $\Delta$ grows with increasing field, and therefore we
performed the experiment in the highest field available.%

The static relaxation rate $\Delta$ is related to lattice deformation via the
muon coupling to its neighboring spins and the system's susceptibility. To
demonstrate this relation let's assume for simplicity that the muon is coupled
only to one electronic spin $\mathbf{S}$ (extension to multiple couplings is
trivial). In this case the magnetic field experienced by the muon is a sum of
the external field $\mathbf{H}_{TF}$ and the field from the neighboring
electron $\mathbf{H}_{int}=g\mu_{B}\mathbf{AS}$ where $g$ is the g factor, and
$\mu_{B}$ is the Bohr magneton. Here $\mathbf{A=A(r)}$ is the coupling between
the muon and electron spin, which we assume depends on the distance between
them, although at present the origin of this coupling is unknown. In a mean
field approximation, $\mathbf{S\rightarrow}$ $\left\langle \mathbf{S}%
\right\rangle =\chi\mathbf{H}_{TF}/g\mu_{B}$. Therefore, the muon experiences
a magnetic field $\mathbf{B}=(1+\mathbf{A\chi})\mathbf{H}_{TF}$. Assuming that
the susceptibility and the couplings are isotropic, the time evolution of a
muon spin due to static field inhomogeneity is given by:
\begin{equation}
P_{static}(t)=\int P_{0}\cos[\gamma_{\mu}(1+A\chi)H_{TF}t]\rho(A)dA.
\label{Rlx_Theory}%
\end{equation}
From Eqs.~(\ref{polarization_root}) and (\ref{Rlx_Theory}) we find that
$\rho(A)=f(\delta A/\{2\pi|A|\})/(2|A|)$ where the effective width of the
distribution, $\delta A$, is given by $\delta A=\left\vert \Delta/(\chi
\gamma_{\mu}H_{TF})\right\vert $ and $f$ can be expressed in terms of
trigonometric and Fresnel functions \cite{Thesis}.

If the ratio between $\Delta$ and $\chi$ remains constant as the temperature
changes, the distribution of couplings $\rho(A)$ is temperature independent.
Unlike most magnets, this is not the case for YMO. In Fig.~\ref{T22p5} we also
present on the lower part of the abscissa the susceptibility measured at 6~kG
with the temperature as an implicit parameter. We see that $\Delta$ is not
proportional to and not even a linear function of $\chi$. We thus conclude
that the distribution of coupling constants broadens extremely fast upon
cooling below 40~K due to random lattice distortions similar to those seen in
the NMR work at higher temperatures~\cite{KerenPRL}.

The local lattice distortions or strains discussed above have been elucidated
from magnetic probes ($\mu$SR and NMR) in a finite field. At first sight these
contradict previously published diffraction data where no structural
deviations from the $Fd\bar{3}m$ space group were
observed~\cite{Raju,Blacklock}. More recent neutron powder diffraction results
in finite applied fields also reveal no unusual thermal dependence on the
oxide pyrochlore structure~\cite{Gardner}. However, Booth and coworkers did
find a $5$\% distribution in Mo-Mo bond lengths in zero field. This quandary,
the apparent conflict between the data sets, required revisiting the theory of
magneto-elastic deformation of the pyrochlore lattice. We have tried to do
this through computer simulations where we posed the questions: (I) Can the
computer find a classical ground state with a lower energy than that suggested
theoretically? (II) How good are the theoretical approximations? and (III)
Since experiments are done at finite temperature, what is the effect of
thermal fluctuations on spin and spatial correlations?

\begin{figure}
[ptb]
\begin{center}
\includegraphics[
natheight=7.619000in,
natwidth=9.694500in,
height=2.4267in,
width=3.0822in
]%
{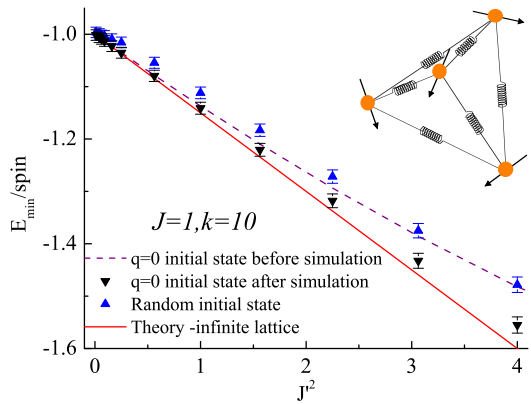}%
\caption{The minimum energy per spin obtained by the simulation program as
described in the text. In the inset we depict a tetrahedral unit of the
pyrochlore lattice with its spins connected by springs to enable visualization
of the magnetoelastic term.}%
\label{EvsJtagsqrc}%
\end{center}
\end{figure}

The Hamiltonian we simulate is the Heisenberg antiferromagnet with a
magneto-elastic term as illustrated in the inset of Fig. \ref{EvsJtagsqrc} for
a single tetrahedon. The rigid bonds between atomic positions of length $R$
are replaced by springs whose elastic energy is zero when $R=1$. This
Hamiltonian is given by
\begin{equation}
\mathcal{H}=J\sum_{i>j}\mathbf{S}_{i}\cdot\mathbf{S}_{j}+J^{\prime}\sum
_{i>j}(\mathbf{S}_{i}\cdot\mathbf{S}_{j}){\delta r_{ij}}+\frac{k}{2}\sum
_{i>j}(\delta r_{ij})^{2} \label{hamil}%
\end{equation}
where $\delta r_{ij}=\left\vert \mathbf{r}_{i}-\mathbf{r}_{j}\right\vert -1$
is the change in bond length between spins $i$ and $j$, $J^{\prime}$ is the
derivative of the exchange $J$ relative to bond length, and $k$ is an elastic
constant. The minimum energy of $\mathcal{H}$ per spin is given by
\begin{equation}
\frac{E_{\min}}{N}=-J-\frac{3J^{\prime2}}{2k} \label{Emin}%
\end{equation}
where $N$ is the number of spins \cite{OlegPRB}. This energy is achieved in
two $q=0$ states with long range displacement and spin order dominated by
E$_{g}$ or E$_{u}$ phonons \cite{OlegPRB}. For both phonons there are 4 short
bonds of length $1-J^{\prime}/k$, and two long bonds of length $\sqrt
{1+2J^{\prime}/k+5(J^{\prime}/k)^{2}}$ on each tetrahedra. The difference
between the phonons is the arrangement of the short and long bonds on the
lattice. In our simulations we use the $q=0$ state with $E_{u}$ phonon where
all tetrahedra of the same orientation on the lattice distort in the same way.

The simulations are performed on a pyrochlore lattice of 8788 spins. In order
to allow for lattice deformation, yet maintain the essence of the spin
frustration, the boundary conditions are periodic for the spins and open for
the coordinates \cite{Simulations}. The energy minimization was done using a
zero temperature Metropolis algorithm which stopped when the change in the
simulated system's energy was below the machine precision of eight digits. The
value of $J^{\prime}$ is variable while $J=1$ and $k=10$ for all runs. Further
details concerning the simulation can be found in Ref.~\cite{Thesis}. The
error bars are determined from an evaluation of the finite size effects: the
energy of the $q=0$ state was calculated using the simulation for different
lattice sizes, and seen to converge to the theoretical value as the lattice
size goes to infinity. Multiple runs of the simulation with the same initial
conditions showed other computational errors to be negligible. It should be
noted that despite the fact that Eq.~\ref{hamil} is valid only for small
$\delta r_{ij}$, in the simulations $\delta r_{ij}$ can assume any number.

In Fig.~\ref{EvsJtagsqrc} we present the energy per spin as obtained for 4
different calculations/simulations: (1) and (2) The energy of the $q=0$ state
given by the computer program before and after energy minimization
respectively. (3) The final energy of the system after energy minimization of
a disordered structure; neither the distortions nor spins entered an ordered
state in this case. (4) Eq.~\ref{Emin} adapted to a finite lattice.

At low values of $J^{\prime2}/(Jk)$ all four cases agree with each other as
expected. However, for high $J^{\prime2}/(Jk)$ the simulation depended
strongly on the initial simulation state, namely, the ground state manifold
for the Hamiltonian is non-ergodic and upon cooling the system can enter one
of many metastable states. The discrepancy between cases (1) and (2) stems
from large values of $\delta r_{ij}$ in the simulation that are not consistent
with the theory. To summarize the simulations: (I) the computer is unable to
find a better state than the $q=0$ one; (II) the theory is valid in the range
$J^{\prime2}/k\lesssim0.1J$; and (III) in this range the disordered states are
very close in energy to the $q=0$ state. This proximity in energy between
ordered and disordered states means that low temperatures are needed to
isolate the $q=0$ state. Eq.~\ref{Emin} suggests that this energy scale is set
by $J^{\prime2}/k$.%

\begin{figure}
[ptb]
\begin{center}
\includegraphics[
natheight=8.454400in,
natwidth=10.810200in,
height=2.4025in,
width=3.0666in
]%
{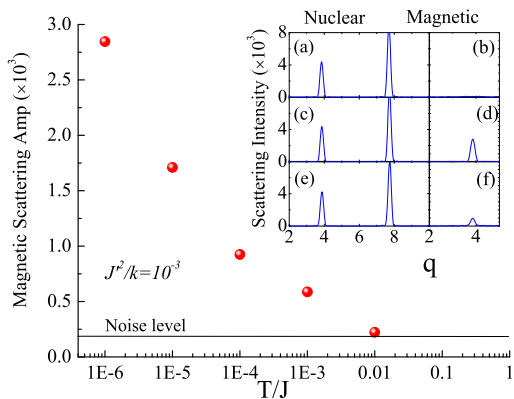}%
\caption{Nuclear and magnetic computer-simulated neutron scattering in the 111
direction. Inset: (a) Nuclear scattering from a perfect pyrochlore lattice.
(b) Magnetic scattering from one of the ground states of the Heisenberg
Hamiltonian (without magneto-elastic term) on this lattice. (c)-(f) Nuclear
and magnetic scattering from the states obtained by a computer simulation of
the Heisenberg Hamiltonian with the magneto-elastic term with $k=10$ and
$J^{\prime}=0.1$ (Eq.~\ref{hamil}) at different temperatures (note the scale
difference). In (c) and (d) $T=10^{-6}J$ and in (e) and (f) $T=10^{-4}J$. Main
figure: magnetic scattering amplitude as a function of temprature.}%
\label{Scattering}%
\end{center}
\end{figure}

Next we examine the temperature dependence of the correlations by simulating a
slow reverse annealing process starting from the $q=0$ state. In order to
follow the displacement and spin correlations we performed computerized
magnetic and nuclear neutron scattering, according to Ref. \cite{squires}, in
the 111 direction. The numerical scatterings of several simulation states are
shown in the inset of Fig. \ref{Scattering}. In panels (a) and (b) we present
a reference scattering at zero temperature from the perfect pyrochlore lattice
with Heisenberg spins and no magneto-elastic term. In panel (a) we see two
peaks which are the trademark of the perfect pyrochlore lattice where the
distance between two planes of the same kind (kagome or triangular) is
$2\sqrt{2/3}R$ and between two different planes (kagome-triangular) is
$\sqrt{2/3}R$. In panel (b) we see the lack of any spin correlations as
expected from the Heisenberg ground states. In panels (c) and (d) we show a
similar experiment, but this time at finite temperature ($T=10^{-6}J$) and
with a magneto-elastic term. Under these conditions long range magnetic order
is found. However, no effect is detected in the nuclear scattering within our
resolution. Panels (e) and (f) are the same as (c) and (d) but at a\ higher
temperature. This reduces both the nuclear and magnetic scattering due to
thermal vibrations, but it affects the magnetic scattering to a greater
extent. This magnetic scattering amplitude is plotted in Fig.~\ref{Scattering}
as a function of temperature. Clearly it decreases as the temperature
increases and is not measurable at $T>J^{\prime2}/k$. This is the main result
of the simulation.%

We use our numerical findings to determine the Hamiltonian parameters of YMO
in the framework of the magneto-elastic coupling. We compare the theoretical
change in bond length, which, to first order, is set by $J^{\prime}/(kR)$ to
the $5$\% distribution in Mo-Mo distance observed by Booth \textit{et al.}
\cite{Booth}. We also estimate $J^{\prime2}/k$ by the temperature at which
spin freezing is observed in YMO, namely, $22.5$~K. From these numbers we
conclude that $J^{\prime}\sim0.01$ eV/\AA \ and $k\sim0.1$~eV/\AA $^{2}$.
\ These numbers should be considered only as order of magnitudes; if one would
take the melting temperature as the point at which the magnetic amplitude
drops to $1/e$ of its $T=0$ value, for example, the estimate of both $k$ and
$J^{\prime}$ would increase by a factor of $10$. For comparison in ZnCr$_{2}%
$O$_{4}$, another well studied frustrated Heisenberg spin system with
approximately the same spin-spin bond length, $J^{\prime}=0.04$ eV/\AA
~\cite{KinoJPSJ72} and $k=6.5$~eV/\AA $^{2}$~\cite{Sushkov}. However, the
reason YMO maintains its cubic symmetry at all temperatures is still a puzzle.
One possibility emerging from the simulation is the lack of ergodicity; upon
cooling the system is stuck in a metastable state and does not relax to its
ground state. In this case the bulk cubic symmetry is maintained, but locally
the system is under immense strain and slightly distorts from its ideal cubic structure.

To summarize, the width of the internal field distribution $\Delta$, as
detected by the $\mu$SR, grows upon cooling at a rate which cannot be
explained by the increasing susceptibility alone. Therefore, we conclude that
the width of the distribution of coupling constants $\delta A$ also grows upon
cooling. We attribute this to temperature dependent lattice distortions.
However diffraction techniques show no structural changes and only local
probes have revealed a distribution of bond lengths. To address this
discrepancy we performed numerical simulations of the Heisenberg Hamiltonian
with magneto elastic coupling at finite temperatures. This allowed us to
estimate $J^{\prime}$ and $k$ in Eq.~\ref{hamil} and to suggest that upon
cooling the system is trapped in a metastable state.

We are grateful for helpful discussion with Oleg~Tchernyshyov and to the
machine and instrument groups at Paul Scherrer Institute, Switzerland, whose
outstanding efforts have made these experiments possible. Work at Brookhaven
National Laboratory is supported by the U.S. Department of Energy under
contracts DE-AC02-98CH10886. The authors wish to acknowledge the financial
support of NATO through a collaborative linkage grant.

\end{document}